\documentclass[reprint,twocolumn,superscriptaddress,showpacs,nofootinbib,notitlepage,prl]{revtex4-1}

\usepackage{graphicx}
\usepackage{latexsym,amsmath,amssymb,lmodern,float,url}
\usepackage{natbib}
\usepackage{color}
\usepackage{microtype}
\usepackage{slashed}
\usepackage{multirow}
\usepackage{xcolor}
\usepackage{tikz}
\usetikzlibrary{shapes}

\usepackage[colorlinks=true,backref=false, linktocpage=true,
citecolor=blue,urlcolor=blue,linkcolor=blue,pdfpagemode=UseOutlines]{hyperref}

\hypersetup{%
  bookmarksnumbered=true,
  pdftitle = {},
  pdfsubject = {},
  pdfauthor = {},
  pdfkeywords = {}
}

\def\Tr {\operatorname{{Tr}}}
\def\Re {\operatorname{{Re}}}

\newcommand{\zn}{\mathbb{Z}_N}

\newcommand{\ds}{S({1080})}

\definecolor{mycolor}{RGB}{22,139,22}
\newcommand{\blacksq}{\raisebox{0.5pt}{\tikz{\node[fill,scale=0.4,regular polygon, regular polygon sides=4,fill=black!20!black](){};}}}
\newcommand{\rcir}{\raisebox{0.5pt}{\tikz{\node[fill,scale=0.4,circle,fill=red](){};}}}
\newcommand{\bltri}{\raisebox{0.7pt}{\tikz{\node[fill,scale=0.3,regular polygon, regular polygon sides=3,fill=blue!10!blue,rotate=0](){};}}}
\newcommand{\blacktri}{\raisebox{0.5pt}{\tikz{\node[draw,scale=0.4,regular polygon, regular polygon sides=3,](){};}}}
\newcommand{\rbcir}{\raisebox{0.5pt}{\tikz{\node[draw,scale=0.4,circle,draw=red](){};}}}
\newcommand{\bluesq}{\raisebox{0.7pt}{\tikz{\node[draw,scale=0.4,regular polygon, regular polygon sides=4,draw=blue,rotate=0](){};}}}
\newcommand{\bluefsq}{\raisebox{0.7pt}{\tikz{\node[draw,scale=0.5,regular polygon, regular polygon sides=4,draw=blue,fill=blue,rotate=0](){};}}}
\newcommand{\redfcir}{\raisebox{0.7pt}{\tikz{\node[draw,scale=0.4,circle,draw=red,fill=red,rotate=0](){};}}}
\newcommand{\greenftri}{\raisebox{0.7pt}{\tikz{\node[draw,scale=0.3,regular polygon, regular polygon sides=3,draw=mycolor,fill=mycolor,rotate=180](){};}}}
\begin{document}

%\title{Modified Actions of $\ds$ for Quantum Computers}
\title{Gluon Field Digitization for Quantum Computers}
\author{Andrei Alexandru}
\email{aalexan@gwu.edu}
\affiliation{Department of Physics, The George Washington University, Washington, D.C. 20052, USA}
\affiliation{Department of Physics, University of Maryland, College Park, MD 20742, USA}
\author{Paulo F. Bedaque}
\email{bedaque@umd.edu}
\affiliation{Department of Physics, University of Maryland, College Park, MD 20742, USA}
\author{Siddhartha Harmalkar}
\email{sharmalk@umd.edu}
\affiliation{Department of Physics, University of Maryland, College Park, MD 20742, USA}
\author{Henry Lamm}
\email{hlamm@umd.edu}
\affiliation{Department of Physics, University of Maryland, College Park, MD 20742, USA}
\author{Scott Lawrence}
\email{srl@umd.edu}
\affiliation{Department of Physics, University of Maryland, College Park, MD 20742, USA}
\author{Neill C. Warrington}
\email{ncwarrin@umd.edu}
\affiliation{Department of Physics, University of Maryland, College Park, MD 20742, USA}

\date{\today}
\collaboration{NuQS Collaboration}
%%%%%%%%%%%%%%%%%%%%%%%%%%%%%%%%%%%%%%%%%%%%%%%%%%%%%%%
\begin{abstract}
Simulations of gauge theories on quantum computers require the digitization of continuous field variables. Digitization schemes that uses the minimum amount of qubits are desirable. 
We present a practical scheme for digitizing $SU(3)$ gauge theories via its discrete subgroup $\ds$.  
The $\ds$ standard Wilson action cannot be used since a phase transition occurs as the coupling is decreased,  well before the scaling regime.
We proposed a modified action that allows simulations in the scaling window and carry out classical Monte Carlo calculations down to lattice spacings of order $a\approx 0.08$ fm. We compute a set of observables with sub-percent precision at multiple lattice spacings and show that the continuum extrapolated value agrees with the full $SU(3)$ results. This suggests that this digitization scheme provides sufficient precision for NISQ-era QCD simulations.
\end{abstract}
%%%%%%%%%%%%%%%%%%%%%%%%%%%%%%%%%%%%%%%%%%%%%%%%%%

\maketitle
%\section{Introduction}
Quantum computers can attack problems in physics which appear intractable on classical computers~\cite{Feynman:1981tf}.  Large-scale quantum computers would allow simulations of non-perturbative calculations of real time evolution and finite-density equations of state. For the foreseeable future, though, quantum computers will be limited to tens or hundreds of non-error-corrected qubits with circuit depths less than a thousand gates --- the so-called Noisy Intermediate-Scale Quantum (NISQ) era. QCD simulations on quantum computers -- especially in the NISQ era -- depend upon formulating QCD in an efficient way. 

Fermionic fields like quarks can be easily digitized as qubit registers by encoding their presence or absence in a given state~\cite{Jordan:1928wi,Verstraete:2005pn,Zohar:2018cwb,2016PhRvA..94c0301W}. This is evident from the few existing calculations performed on quantum computers~\cite{Martinez:2016yna,Klco:2018kyo,Lamm:2018siq,Shehab:2019gfn}. The continuous nature of gauge fields preclude such exact digitization. Proposed solutions involve either eliminating the bosonic fields using some model-dependent properties or truncating in occupation number~\cite{Hackett:2018cel,Macridin:2018gdw,Yeter-Aydeniz:2018mix,Klco:2018zqz,Bazavov:2015kka,Zhang:2018ufj,Unmuth-Yockey:2018xak,Unmuth-Yockey:2018ugm,Zache:2018jbt,Raychowdhury:2018osk,Kaplan:2018vnj,Stryker:2018efp,Alexandru:2019ozf}. 

The situation is reminiscent of the pioneering days of lattice field theory when %classical bits were precious 
computer memory was limited
and the cost of storing $SU(3)$ elements was prohibitive. Several attempts were made to replace the continuous gauge fields by a discrete set of values~\cite{Petcher:1980cq,Jacobs:1980mk,Bhanot:1981xp,Grosse:1981bv,Bhanot:1981pj,Lisboa:1982jj,Flyvbjerg:1984dj,Flyvbjerg:1984ji}.  
Quantum computation is presently in a similar situation where  every  qubit comes at a high cost. Representing each gauge link by 9 complex numbers represented using double-precision format requires 1152 qubits.
%valued, double-precision numbers requires 1152 qubits.  
In contrast, the largest ``crystal-like'' discrete subgroup of $SU(3)$, $\ds$, contains 1080 elements and would require only 11 qubits to store each link value.  
%So, for the price of one $SU(3)$ link, one can represent a $4^3$ lattice of $\ds$.   
%\footnote{By crystal-like, we mean that any two nearest neighbors to the identity can be related by an isometry of the matrix group. In the case of S(1080), that symmetry is given by the adjoint action of S(1080) itself.}

Digitization typically reduces the symmetry of the model.  With this reduction, it is not a given that the original model is recovered in the continuum limit as the universality class of the lattice model may change~\cite{Hasenfratz:2001iz,Caracciolo:2001jd,Hasenfratz:2000hd,PhysRevE.57.111,PhysRevE.94.022134,article}. For any discrete group, there is always a finite difference in the action, $\Delta S$, between the field configurations with the two smallest actions, as opposed to continuous groups where no such gap exists.  This 
may lead to  ``freezing'' at some critical $\beta_f$; that is, all field values except the identity (and gauge-equivalents) are exponentially suppressed.  For values of $\beta$ beyond $\beta_f$ the theory with the discrete group differs drastically from the continuous group and is no longer a reasonable approximation.  This is a particular problem for asymptotically-free theories like QCD. The spacetime continuum limit where the lattice spacing $a$ approaches zero is obtained by making $\beta$ large but that is where the continuous and discrete group theories differ. This is not necessarily fatal: realistic lattice calculations are performed on classical computers with a finite $a$ and extrapolated in a controlled manner to $a\rightarrow0$. In these calculations, what is required is that $a$ is smaller than typical hadronic scales (e.g. the size of hadrons). Typically values used in state-of-art calculations are $\mathcal{O}(0.1$ fm). This corresponds $\beta \agt 6$ when using the Wilson action, in the so-called scaling region.  Our goal will be to set up a framework where discrete groups can be used to reproduce $SU(3)$ results in the scaling region,  such that continuum extrapolations can be performed.

There were a number of early studies of the viability of crystal-like discrete subgroups of $U(1)$~\cite{Creutz:1979zg,Creutz:1982dn} and $SU(N)$~\cite{Bhanot:1981xp,Petcher:1980cq,Bhanot:1981pj} gauge theories.  While the discrete subgroups all have freezing transitions for the Wilson action, $\beta_f$ increases with the size of the subgroup.  For $U(1)$, the $\zn$ theories have a $\beta_f$  in the scaling regime (in this case $\beta\agt 1$) for  $N>4$.  $SU(2)$ has only three crystal-like subgroups: the binary tetrahedral, $\mathbb{BT}$, the binary octahedral, $\mathbb{BO}$, and the binary icosohedral, $\mathbb{BI}$.  Using the Metropolis algorithm with 100 measurements separated by 1000 updates, we refined the results of~\cite{Petcher:1980cq}, finding that while $\mathbb{BT}$ has $\beta_f=2.24(8)$, $\mathbb{BO}$ and $\mathbb{BI}$ have $\beta_f=3.26(8)$ and $\beta_f=5.82(8)$ respectively, both deep in the scaling regime $\beta\agt2.2$. Hence, these two last groups can be used in lieu of $SU(2)$ for practical calculations.

The story changes for $SU(3)$.  There are five crystal-like subgroups: $S(60)$, $S(108)$, $S(216)$, $S(648)$, and $\ds$, designated by their number of elements.  For all these, $\beta_f < 6$ , with the largest, $\ds$, being reported to have $\beta_f=3.58(2)$ obtained on a $2^4$ lattice~\cite{Bhanot:1981xp}.  Our own calculations on a~$2^4$ volume with larger statistics for $\ds$ show a slightly larger value $\beta_f=3.935(5)$. In any case, it is evident that the $\ds$ theory with the Wilson action is inadequate to reach the scaling regime. Subsequent work~\cite{Lisboa:1982jj} showed that extending the elements to include the midpoints between elements of $\ds$ was sufficient to push $\beta_f\approx7$. However this requires more bits and sacrifices the group structure.

To overcome these limitations, attempts were made to approximate the $SU(3)$ Wilson action by a 
modified action based on a subgroup~\cite{Edgar:1981dr,Bhanot:1981pj,Creutz:1982dn,Fukugita:1982kk,Horn:1982ef,Flyvbjerg:1984dj,Flyvbjerg:1984ji,Ayala:1989it}, although only in~\cite{Bhanot:1981pj} were Monte Carlo calculations undertaken. 
There, simulations using $S(648)$ with a Wilson action modified by a $|\Tr U_p|^2$ term (equivalent to the trace in the adjoint representation) were performed. Even with this modified action $S(648)$ was inadequate to reach the scaling regime.  Further, it was conjectured, based on small scale simulations and mean-field estimates,
%the value $\beta_f=3.58(2)$ for $\ds$ and mean-field estimates, 
that $\ds$ would also fail to reach the scaling regime with that modified action.
 Calculations with modified actions of $SU(3)$ were also performed to study thermodynamics and reduce lattice spacing errors~\cite{Blum:1994xb,Heller:1995bz,Heller:1995hh,Hasenbusch:2004wu,Hasenbusch:2004yq}.  

%\section{Results}

With this history in mind, we study the viability of $\ds$ with a different action
%differently modified action by performing lattice Monte Carlo simulations.  The full action is given by
\begin{equation}\label{eq:action-mod}
 S=-\sum_p \left(\frac{\beta_0}{3}\Re\Tr U_p +\beta_1\Re\Tr\, U_p^2\right)\,,
\end{equation}
where $U_p \in S(1080)$ indicates a plaquette, and the first term has been normalized such that, for $\beta_1=0$, the action matches the $SU(3)$ Wilson action (with $\beta=\beta_0$).  
%We define two energies, 
%\begin{align}
% \langle E_0\rangle=\frac{1}{V}\frac{\partial \langle S\rangle}{\partial \beta_0}=\frac{1}{V}\langle \Re\Tr U_P\rangle,\nonumber\\ \langle E_1\rangle=\frac{1}{V}\frac{\partial \langle S\rangle}{\partial \beta_1}=\frac{1}{V}\langle \Re\Tr U_P^2\rangle
%\end{align}
Simulations can be performed efficiently by employing precomputed multiplication and trace tables.

\begin{figure}
\begin{center}
 \begin{picture}(250,150)
 \put(0,0){\includegraphics[width=\linewidth]{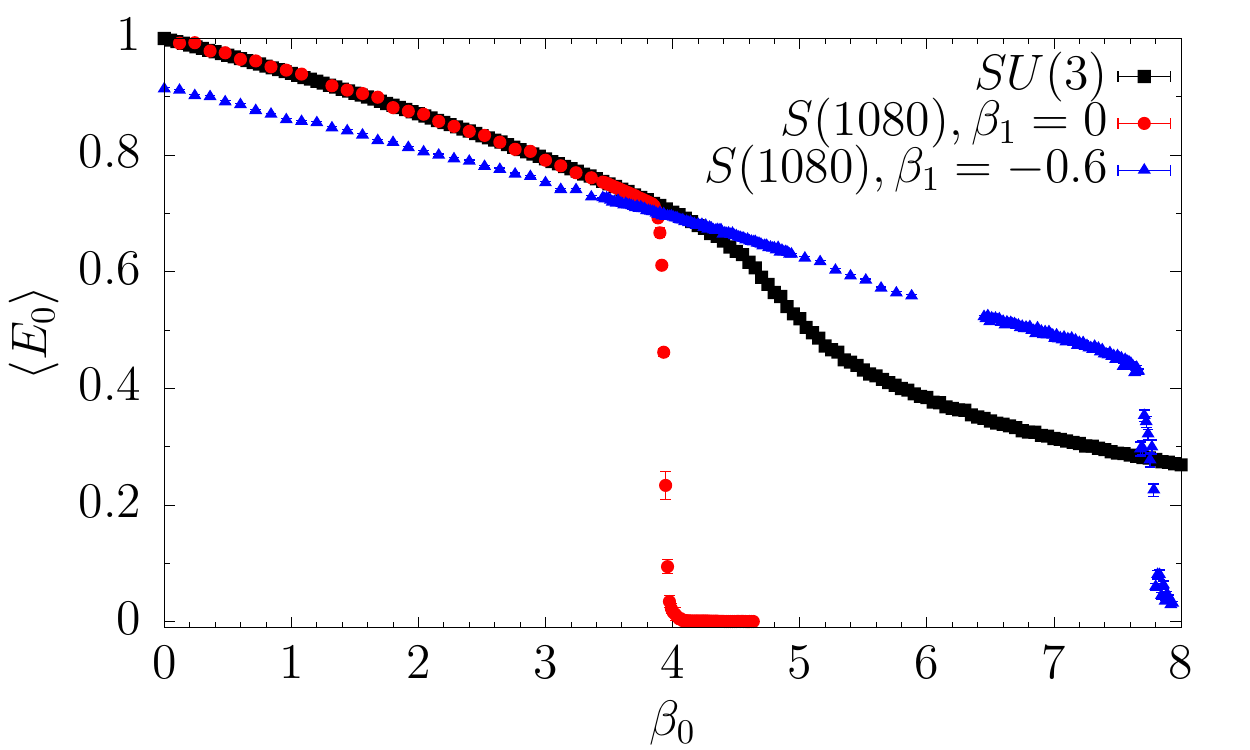}}
  \put(32,25){\includegraphics[width=0.42\linewidth]{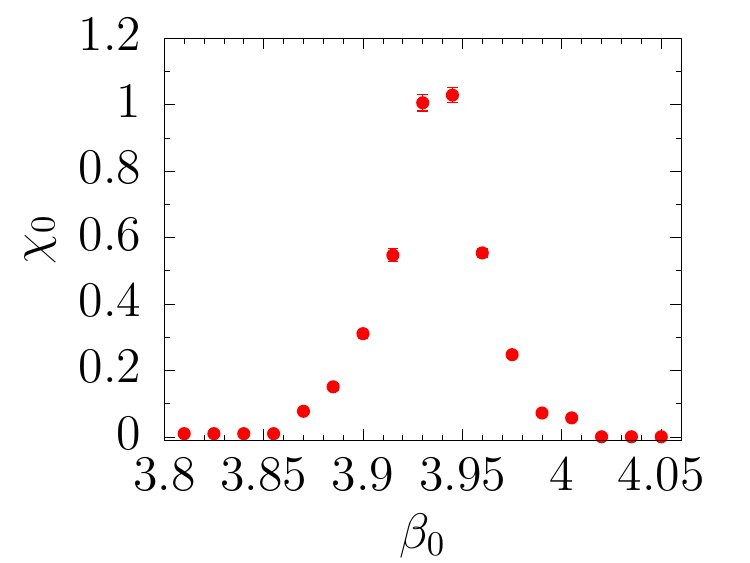}}
\end{picture}
\end{center}
\caption{\label{fig:plaq}Average energy per plaquette, $\langle E_0\rangle = 1-\Re\langle\Tr U_p\rangle /3$, vs $\beta_0$ on $2^4$ lattice for: (\protect\blacksq) $SU(3)$  and $\ds$ with (\protect\rcir) $\beta_1=0$  and with (\protect\bltri) $\beta_1=-0.6$. (inset) $\chi_0$ vs $\beta_0$ for $\beta_1=0$.}
\end{figure}

One could argue heuristically that the action in Eq.~(\ref{eq:action-mod}) will lead to the same continuum limit as, for instance, the Wilson action of $SU(3)$ ($S_{SU(3)} = -\tfrac{\beta_{SU(3)}}{3} \sum_p \Re\Tr U_p$) by noting that the continuum limit in asymptotically-free theories is obtained by setting the coupling to be small ($\beta$ large) and that in this limit only small field fluctuations are important. For small values of the field $S$ and $S_{SU(3)}$ agree as long as $\beta_{SU(3)}$ is a certain linear combination of $\beta_0, \beta_1$. The flaw with this argument is that as $\beta_{SU(3)}$ is increased towards the continuum limit, so do $\beta_0, \beta_1$ and, at some point, the fluctuations become smaller than the separation between the identity and the nearest element to it in $\ds$ and fields freeze. Beyond this point it makes no sense to consider small fluctuations in the $\ds$ theory. This is seen dramatically in Fig.~\ref{fig:plaq} where the average plaquette of the Wilson action $SU(3)$ theory (black) and the $\ds$ action $S$ at $\beta_1=0$ (red) are shown to agree very well until they abruptly diverge when freezing occurs. This agreement is expected since the strong coupling expansion predicts the difference in the average plaquette for $SU(3)$ and $S(1080)$ with $\beta_1=0$ to be of order ${\cal O}(10^{-6}\beta^5)$~\cite{Bhanot:1981xp}. By setting $\beta_1<0$ the gap between the two lowest action values is reduced and freezing occurs only at large $\beta_0$~(see blue curve in Fig.~\ref{fig:plaq}.) 
%Thus, for $\beta_1<0$, action $S$ gets closer to the continuum limit of $S_{SU(3)}$. 
Our proposal is to find a trajectory in the ($\beta_0, \beta_1$) plane that avoids freezing and allows us to get closer to the continuum limit. The idea is that as we move on this trajectory towards larger values of $\beta_0$ we produce configurations with larger correlation lengths, ideally increasing all the way to infinity. We must then check that this action generates the same physics as $SU(3)$.

\begin{figure}
 \begin{center}
  \includegraphics[width=\linewidth]{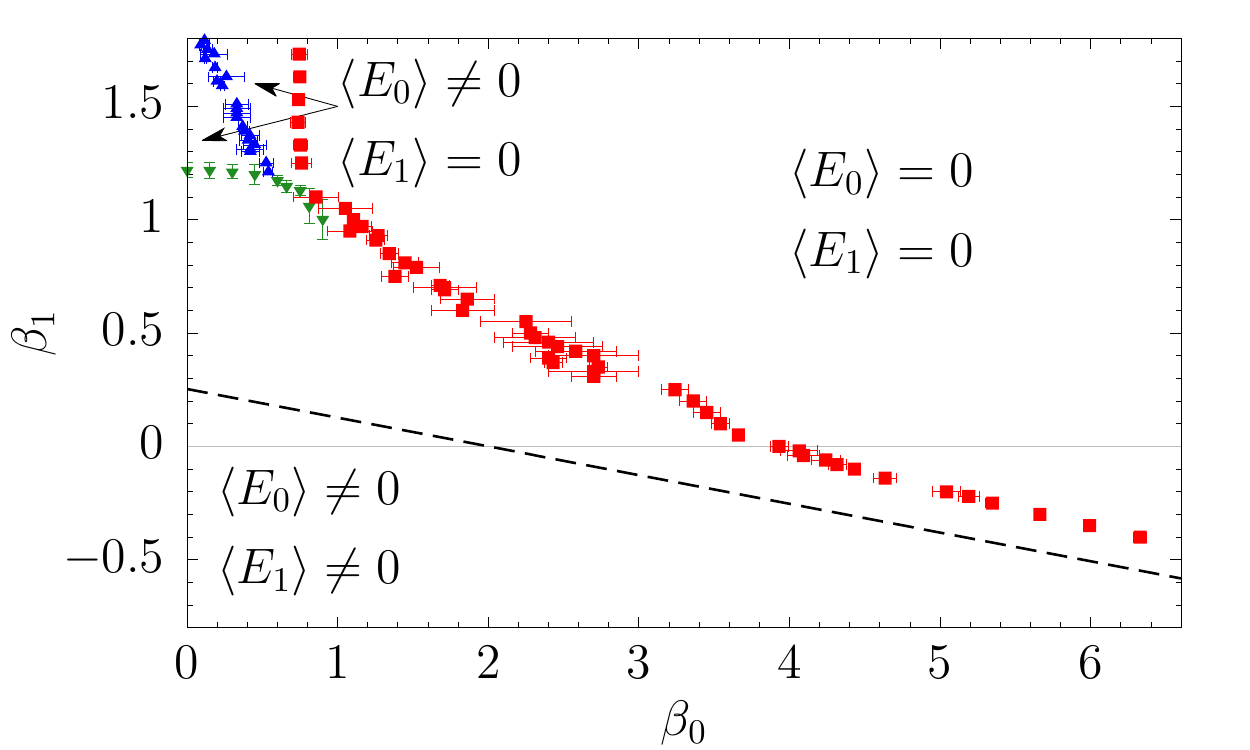}
 \end{center}
\caption{\label{fig:pd}Phase diagram in the $(\beta_0,\beta_1)$ plane for a $2^4$ lattice.  Also shown as a dashed line is the trajectory used for exploring larger lattices.}
\end{figure}

In order to guide our choice of trajectory we roughly mapping out the phase diagram of the $\ds$ theory in the ($\beta_0, \beta_1$) plane to determine where the theory is frozen and not useful to approximate $SU(3)$. The identification of this boundary is complicated because each phase exists as a metastable state throughout the phase diagram.
To deal with this metastability and the associated long autocorrelation times, we used a parallel tempering algorithm similar to~\cite{PhysRevLett.57.2607,B509983H}.  We perform simulations with a set of $\{\beta_{0,i}\}$ for a fixed $\beta_1$.  For every fifth local update, the configuration with $\beta_{0,i}$ is randomly selected and swapped with the ensemble with $\beta_{0,j}$ where $j=i\pm1,i\pm2$ with probability:
\begin{equation}
 P_{ij}(\phi_i,\phi_{j})=\text{min}\left(1,e^{(\beta_{0,i}-\beta_{0,j})(\tilde{S}[\phi_i]-\tilde{S}[\phi_j])}\right).
\end{equation}
%To further reduce the tunneling time between phases, we initialize half of our ensembles each into hot/cold starts.  
With this algorithm, we were able to map out the full $(\beta_0,\beta_1)$ space by searching for peaks in the susceptibilities  $\chi_0=\frac{\partial^2\langle S\rangle}{\partial\beta_0^2}$ and/or $\chi_1=\frac{\partial^2\langle S\rangle}{\partial\beta_1^2}$ (see Fig.~(\ref{fig:pd})).  Besides the freezing transition (shown in red), there are additional transitions 
shown in the upper left corner of the phase diagram delineating regions where the dynamics is partially frozen down to a subset of the group elements. 
This rich structure is qualitatively similar to the one found in the $SU(3)$ theory but it is of little concern to us. 
We focus instead on the lower ($\beta_1 <0$) region of the phase diagram.
At $\beta_1=0$ we measure $\beta_c=3.935(5)$, a larger value
than the $\beta_c=3.58(2)$ found with smaller statistics in~\cite{Bhanot:1981xp}. 

With the freezing transition mapped out, we can propose a way to approach the continuum limit by performing simulations along a trajectory that avoids the frozen phase.  
We choose the trajectory
\begin{equation}
\label{eq:traj}
 \beta_1=-0.1267\beta_0+0.253,
\end{equation} shown as a dotted line in Fig.~\ref{fig:pd}.
A pure gauge theory (without matter) can predict only ratios of observables. So, one observable should be used to set the scale, that is, to determine the dimensionful quantity $a$ at a fixed $\beta_0, \beta_1$. After this, the theory can make predictions for all other observables.  In order to quantify the approach to the continuum and compare it with the expected continuum result we then need a minimum of two observables: one to set the scale and another to be predicted. We choose the pseudocritical temperature $T_c$ for deconfinement and the scale $t_0$ defined by the Wilson flow~\cite{Luscher:2010iy}. 

We determine $T_c$ by looking at the distribution of values of the Polyakov loop $P$. In the confined phase they concentrate around zero, while at higher temperatures cluster around the vertices of a triangle. We denote by $w_c, w_3$ the number of configurations in which $P$ lies near zero or near one of the vertices of the triangle. Following the procedure outlined in~\cite{Francis:2015lha}, we label as ``center'' configurations the ones that lie inside an equilateral triangle centered at
the origin and rotated such that one side is perpendicular to the positive real axis. The separatrix is adjusted to the minimum of the histogram of the $\Re\Tr P$. This position is determined from simulations close to the transition point. There is some ambiguity in this definition but we include this variability into our error budget. We then look at the quantity  $s(\beta_0, \beta_1)=\frac{3w_c-w_3}{3w_c+w_3}$: it is $+1$ deep in the confined phase and reaches -1 deep in the deconfined phase. When $s=0$, the theory is tuned to $T_c$. This definition of $T_c$ has been shown to have finite-volume effects that scale exponentially in the spatial volume for $SU(3)$, unlike the peak of $\chi_0$ which exhibits power law volume corrections~\cite{Francis:2015lha}.  We found that for $N_t=4$, the difference in $(\beta_0^c,\beta_1^c)$ for $N_s=12,16$ was negligible, confirming the same behavior in $\ds$. Therefore, we assumed $N_s=3N_t$ has negligible finite-volume effects at larger $N_t$ and use this volume to compute $T_c$. For each set of parameters we collected ${\cal O}(10^6)$ measurements separated by 10 sweeps. To perform a sweep, we visit each link and update it using a multi-hit Metropolis step. 

For $N_t=4,6,8$ we scan ($\beta_0, \beta_1$) along the trajectory Eq.~(\ref{eq:traj}) to find ($\beta^c_0, \beta^c_1$) for which $s=0$. The values obtained are listed on Table~\ref{tab:vals}. For each of these the inverse physical temperature $1/T = N_t a(\beta_0^c, \beta_1^c) = 1/T_c$ is the same.

\begin{table}[b]
\caption{\label{tab:vals} Wilson flow parameters $\sqrt{t_0}/a$ and $\sqrt{t_{0.2}}/a$ found on lattices of size $(3N_t)^4$ along our trajectory where $N_t$ is temporal size used to determine $T_c$.  In the last two columns, the first error is statistical, and the second is from the separatrix.}
\begin{center}
\begin{tabular}
{c c | c c | c c}
\hline\hline
$N_{t}$ & $N_s$& $\beta^c_0$ & $\beta^c_1$ & $\sqrt{t_0}/a$ & $\sqrt{t_{0.2}}/a$\\
\hline
4 & 12 & 9.154(2) & -0.9061(3) &1.016(3)(3)&0.8316(12)(20)\\
6 & 18 & 12.795(9) & -1.3673(11) &1.508(3)(5) &1.2068(18)(42)\\
8 & 24& 19.61(4) & -2.231(5) &2.000(4)(8)   &1.595(3)(6)\\
\hline\hline
\end{tabular}
\end{center}
\end{table}

We compute the scale $t_0$ defined by the Wilson flow for these sets of $(\beta^c_0, \beta^c_1)$ parameters. We first generate
configurations on lattices of size $(3N_t)^4$ where $N_t$ is the temporal size used to determine $T_c$.  For each ensemble we generate 200 configurations, separated by 1000 sweeps.
The configurations generated are represented as $SU(3)$ matrices and used as initial conditions for performing the Wilson flow~\cite{Luscher:2010iy}: 
\begin{align}
 \dot{V}_t(x,\mu)=-\frac{1}{\beta_0}&\left(\partial_{x,\mu}S_W[V_t]\right)V_t(x,\mu),\nonumber\\V_t(x,\mu)|_{t=0}&=U(x,\mu)
\end{align}
where $S_W[V_t]$ is the Wilson action of $SU(3)$ fields $V_t$ at some Wilson-flow time $t$.  
%This flow corresponds to an infinitesimal stout smearing of the gauge fields.  
%Although $V_t$ are valued on $SU(3)$ instead of $\ds$, this is a well-defined observable for $\ds$ elements since the flow commutes with $\ds$ gauge transformations.  
Using the flow, we define two observables $t_{X=0.2}$ and $t_{X=0.3}$ implicitly by the expression
\begin{equation}
 \left(t^2\langle E\rangle\right)_{t=t_X}=X
\end{equation}
where $X=0.2,0.3$, and
% $E=\frac{1}{V}\sum_P\Re\Tr U_P$.   
$E$ is the lattice clover definition of the energy density.
%In this work we have used the Wilson term, $E=\frac{1}{V}\sum_P\Re\Tr U_P$.  Other choices of $E$ have been investigated in \cite{Francis:2015lha} by using Wilson-improved and clover operators.  
Following the convention, $t_{X=0.3}$ is called $t_{0}$.  Both of these observables have been measured precisely for $SU(3)$ pure gauge theory, allowing for comparison. By also measuring $t_{0.2}$, we probe higher energy scales where larger discrepancies between $\ds$ and $SU(3)$ should be found.  Our results are found in Table~\ref{tab:vals}. 

\begin{figure}
 \begin{center}
  \includegraphics[width=\linewidth]{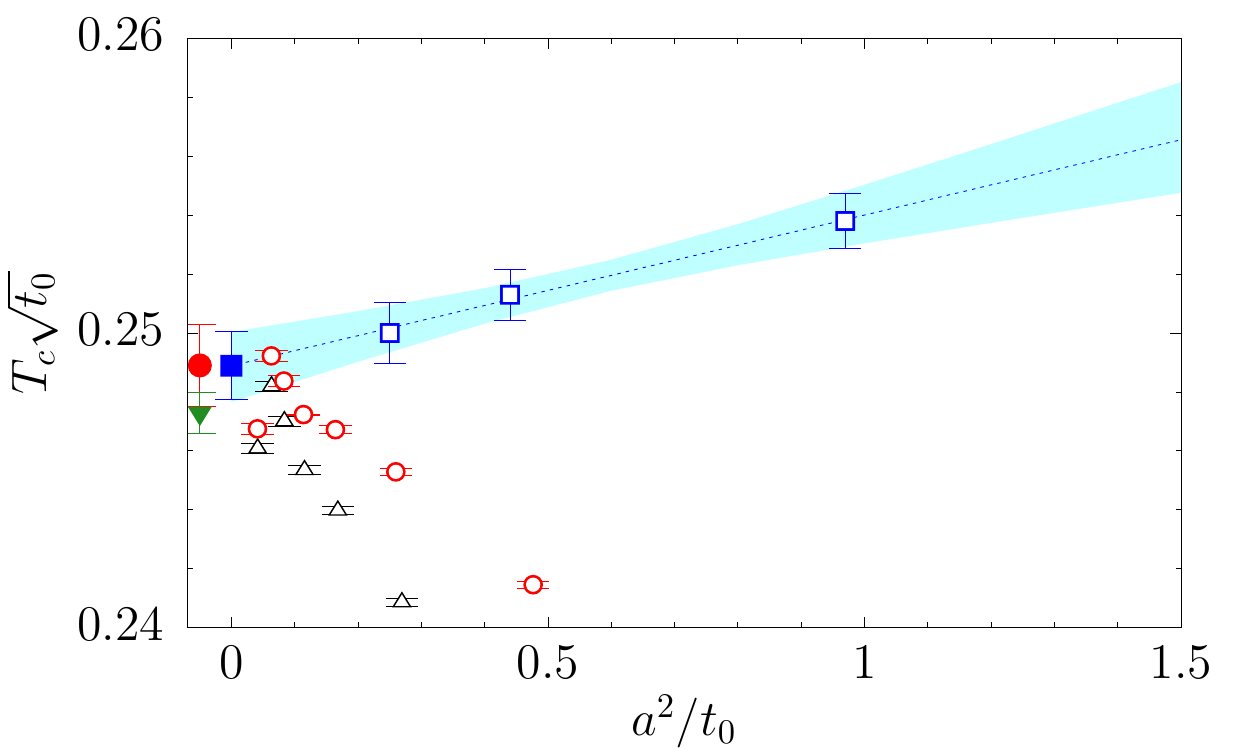}
    \includegraphics[width=\linewidth]{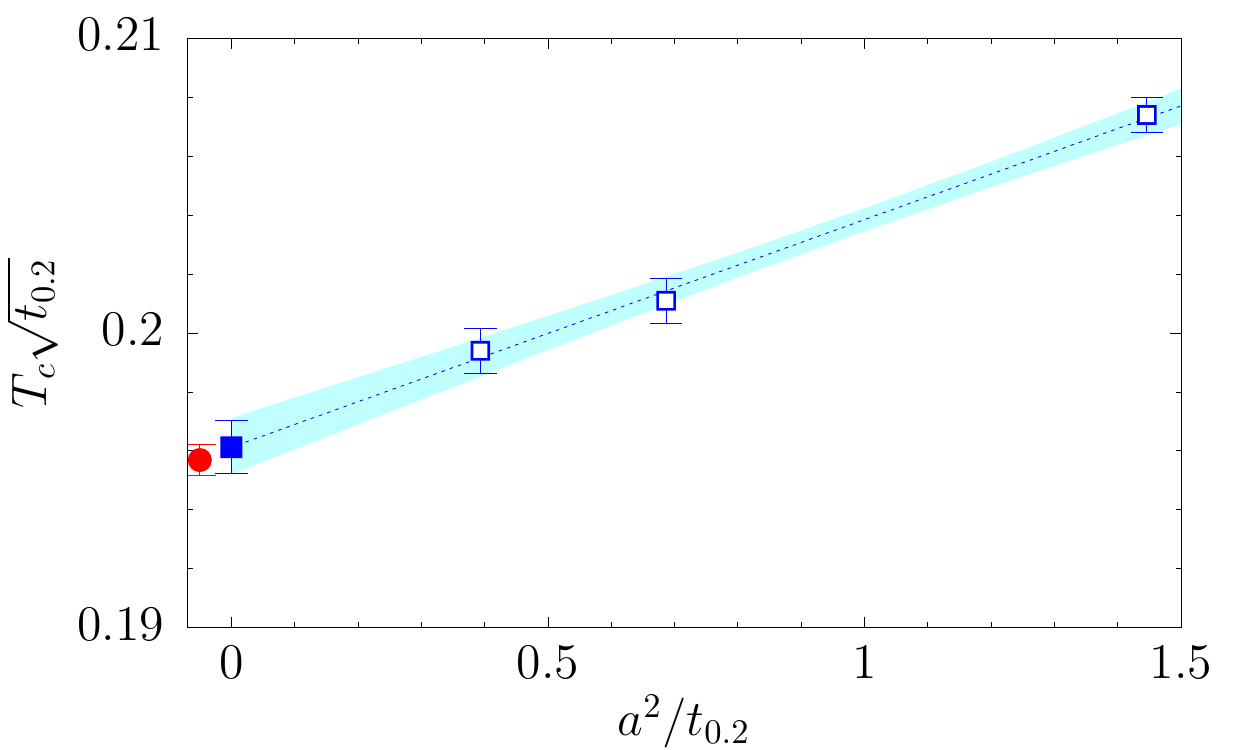}
 \end{center}
\caption{\label{fig:tct0}(top) $T_c\sqrt{t_{0}}$ vs $a^2/t_0$.  Our results (\protect\bluesq) compared to $SU(3)$ results from \cite{Francis:2015lha} using Wilson~(\protect\blacktri) and Wilson-improved~(\protect\rbcir) energies are reproduced for comparison. Our extrapolated value (\protect\bluefsq) are compared to $SU(3)$ results of~\cite{Francis:2015lha} (\protect\redfcir) and~\cite{Kitazawa:2016dsl} (\protect\greenftri). (bottom) $T_c\sqrt{t_{0.2}}$ vs $a^2/t_{0.2}$ and the extrapolation compared with
the results of \cite{Asakawa:2015vta} with same symbols.}
\end{figure}

In the absence of discretization effects, the value of $t_0$ in physical units should be the same on all our lattices. We demonstrate that the variation of $t_0$ as we approach the continuum is mild and the extrapolated value agrees with the full $SU(3)$ result.  With our data, it is possible to construct a dimensionless quantity, $T_c\sqrt{t_X}$ which can be compared to those of $SU(3)$ at both finite lattice spacing $a$ and by extrapolating to the continuum.  Using a linear extrapolation, we compute a continuum value of $T_c\sqrt{t_{0}}=0.2489(11)$ which is in agreement with $T_c\sqrt{t_{0}}= 0.2489(14)$~\cite{Francis:2015lha} and $T_c\sqrt{t_{0}}= 0.2473(7)$~\cite{Kitazawa:2016dsl} computed in full $SU(3)$. Similarly, our extrapolated value of $\frac{\sqrt{t_0}}{\sqrt{t_{0.2}}}=0.1269(6)$ is in good agreement with the value of $0.1264(4)$ computed for~$SU(3)$~\cite{Asakawa:2015vta}.
 Our results for $T_c\sqrt{t_0}$ are compared to~\cite{Francis:2015lha} in Fig.~\ref{fig:tct0}.  It is interesting to note that the $\mathcal{O}(a^2)$ corrections appear milder for the modified action used here compared to the Wilson action of the $SU(3)$.  This feature of modified actions has been observed previously in $SU(3)$~\cite{Blum:1994xb,Heller:1995bz,Heller:1995hh,Hasenbusch:2004wu,Hasenbusch:2004yq}, which suggest further benefits of using this action for quantum simulations. 

Assuming our $S(1080)$ action has the same continuum limit as $SU(3)$, it is possible derive physical values for $t_0$ and a lattice spacing $a$ for each of our ensembles. Using the $SU(3)$ relations of~\cite{Asakawa:2015vta,Kitazawa:2016dsl} between $w_{0.4}$, $T_c$, $\Lambda_{SU(3)}$ and $r_0=0.49(4)$ fm, we obtain $\sqrt{t_0}\approx0.16(2)$~fm where our error is dominated by $r_0$, agreeing with $\sqrt{t_0}=0.1638(10)$~fm determined in $SU(3)$~\cite{Sommer:2014mea}. With this, our $24^4$ lattice has $a\approx0.08$~fm or 2.5~GeV$^{-1}$.  This suggests it would be possible to extract and compare glueball states~\cite{Morningstar:1999rf,Chen:2005mg} or quenched calculations of hadron masses~\cite{Aoki:2002fd} to $SU(3)$ values with sub-percent precision.

%\section{Discussion and Conclusion}
In this work, we have presented results for the discrete gauge group $\ds$ using a modified Wilson action.  After mapping the entire phase diagram of the new action, we found a parameter trajectory that avoids the freezing transition, allowing for calculations 
with lattice spacing $a\approx 0.08$ fm. We have shown  that this action is capable of reproducing the physics of $SU(3)$ below 2.5 GeV$^{-1}$ by measuring $T_c\sqrt{t_{0}}=0.2489(11)$ and $\frac{\sqrt{t_0}}{\sqrt{t_{0.2}}}=0.1269(6)$, which agree to remarkable precision with the full group.

The qubit savings from using $\ds$ instead of $SU(3)$ are dramatic.  In the NISQ era, where small lattice sizes and noisy gates will likely dominate the error, the parameters used in this paper should be a sufficient approximation of $SU(3)$.  However, if gluon actions are required at $a<0.08$ fm, the action in~Eq.~(\ref{eq:action-mod}) on the trajectory specified by~Eq.~(\ref{eq:traj}) may be insufficient since $a$ on this trajectory seems to have a minimum.  Another trajectory nearer to the freezing transition may provide a smaller lattice spacing. Future work is required to determine whether we can generate arbitrarily small $a$ on a different trajectory with our action or if additional terms are required. Furthermore, questions of how well other observables like hadronic spectra are reproduced and the effect of including of fermions are left for future studies.  Given that circuit depth is of concern in the NISQ era, a broad search in modified action space should be undertaken with an eye toward terms that require few quantum gates while still efficiently reaching the continuum limit of $SU(3)$.  Quantitative comparisons of these actions will require constructing the quantum gates for $\ds$~\cite{Lamm:2019bik}.

\begin{acknowledgments}
%The authors thank Neill Warrington for valuable discussions on parallel tempering methods.
 The authors thank Y. Yamauchi and Y. Cai for insight into quantum gates.  A. A. is supported in part by the National Science Foundation CAREER grant PHY-1151648 and by U.S. Department of Energy under Contract No. DE-FG02-95ER-40907. A.A. acknowledges the hospitality of the University of Maryland where part of this work was performed.  P. B., S.H., H. L., and S. L. are supported by the U.S. Department of Energy under Contract No.~DE-FG02-93ER-40762. H. L. acknowledges the hospitality of U.W. where parts of this work were produced.
\end{acknowledgments}
\bibliographystyle{apsrev4-1}
\bibliography{wise2}
\end{document}